\documentclass{aa}
\usepackage{psfig}
\usepackage{color}
\usepackage{natbib}

\def\h2{H{\small II}}

\newcounter{qub}
\begin{document}

\title{Two extremely metal-poor emission-line galaxies in the Sloan Digital
Sky Survey\thanks{Based on observations
collected at the European Southern Observatory, Chile, ESO program 
76.B-0739.}
}

\author{Y. I. Izotov \inst{1}
\and P.\ Papaderos \inst{2}
\and N. G.\ Guseva \inst{1}
\and K. J.\ Fricke \inst{2}
\and T. X.\ Thuan\inst{3}}
\offprints{Y. I. Izotov, izotov@mao.kiev.ua}
\institute{      Main Astronomical Observatory,
                     Ukrainian National Academy of Sciences,
                     Zabolotnoho 27, Kyiv 03680,  Ukraine
\and
                     Institute for Astrophysics, Friedrich-Hund-Platz 1,
                     37077 G\"ottingen, Germany
\and
                     Astronomy Department, University of Virginia,
                     Charlottesville, VA 22903, USA
}

\date{Received \hskip 2cm; Accepted}

\abstract{We present spectroscopic observations with the 3.6m ESO telescope
of two emission-line galaxies, J2104$-$0035 and J0113$+$0052, 
selected from the Data Release 4 (DR4) of the Sloan Digital Sky Survey
(SDSS). 
From our data we determine the oxygen abundance of these systems
to be respectively 12 + log O/H = 7.26 $\pm$ 0.03 and 7.17 $\pm$ 0.09, 
making them the two most metal-deficient galaxies found thus far 
in the SDSS and placing them among the five most metal-deficient
emission-line galaxies ever discovered. 
Their oxygen abundances are close to those of the two most
metal-deficient emission-line galaxies known, SBS 0335--052W with 
12 + log O/H = 7.12 $\pm$ 0.03 and I Zw 18 with 12 + log O/H = 7.17 $\pm$ 0.01.
}
\titlerunning{Two most metal-poor emission-line galaxies found in the Sloan Digital Sky Survey}
\maketitle


\section{Introduction}

Extremely metal-deficient emission-line galaxies at low redshifts are the 
most promising young galaxy candidates in the local Universe
\citep{G03,IT04b}. The studies of their nearly pristine interstellar 
medium (ISM) can also 
shed light on the properties of the primordial ISM at the 
time of galaxy formation.     
 It appears now that even the most metal-deficient
galaxies in the local Universe formed from matter which was already 
pre-enriched by a previous star formation episode, e.g. by 
Population III stars
\citep{T05}. It is thus quite important to establish firmly 
the level of this pre-enrichment
by searching for the most metal-deficient emission-line galaxies. 
Because they have not undergone much chemical evolution, these galaxies
are also the best objects for the determination of the primordial He 
abundance and for constraining cosmological models \citep[e.g. ][]{IT04a}.

However, extremely metal-deficient emission-line galaxies are very rare. 
Many surveys have been carried out 
to search for such galaxies without a large success. For more than three
decades,
one of the first blue compact dwarf (BCD) galaxies discovered, I Zw 18 
\citep{SS70} continued to hold the record as the most metal-deficient 
emission-line galaxy known, with
an oxygen abundance 12 + log O/H = 7.17 $\pm$ 0.01 in its 
northwestern component 
and 7.22 $\pm$ 0.02 in its southeastern component \citep{TI05}. 
Only very recently, has I Zw 18 been displaced by the BCD 
SBS 0335--052W. This galaxy , with an oxygen abundance 
12 + log O/H = 7.12 $\pm$ 0.03, is now the emission-line galaxy 
with the lowest metallicity known \citep{I05}.

\begin{figure*}[t]
\hspace*{-0.0cm}\psfig{figure=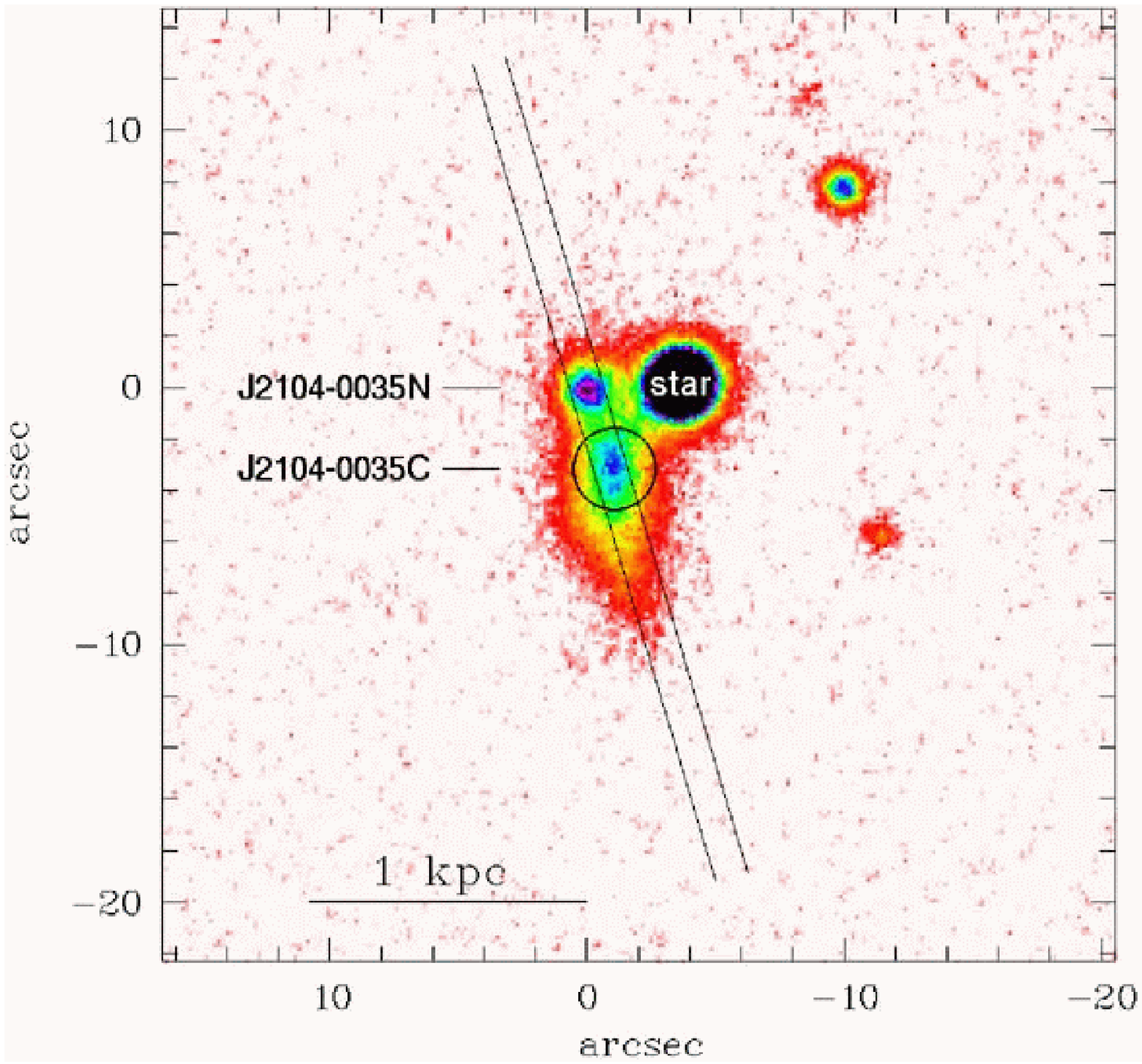,angle=0,width=8.0cm,clip=}
\hspace*{-0.0cm}\psfig{figure=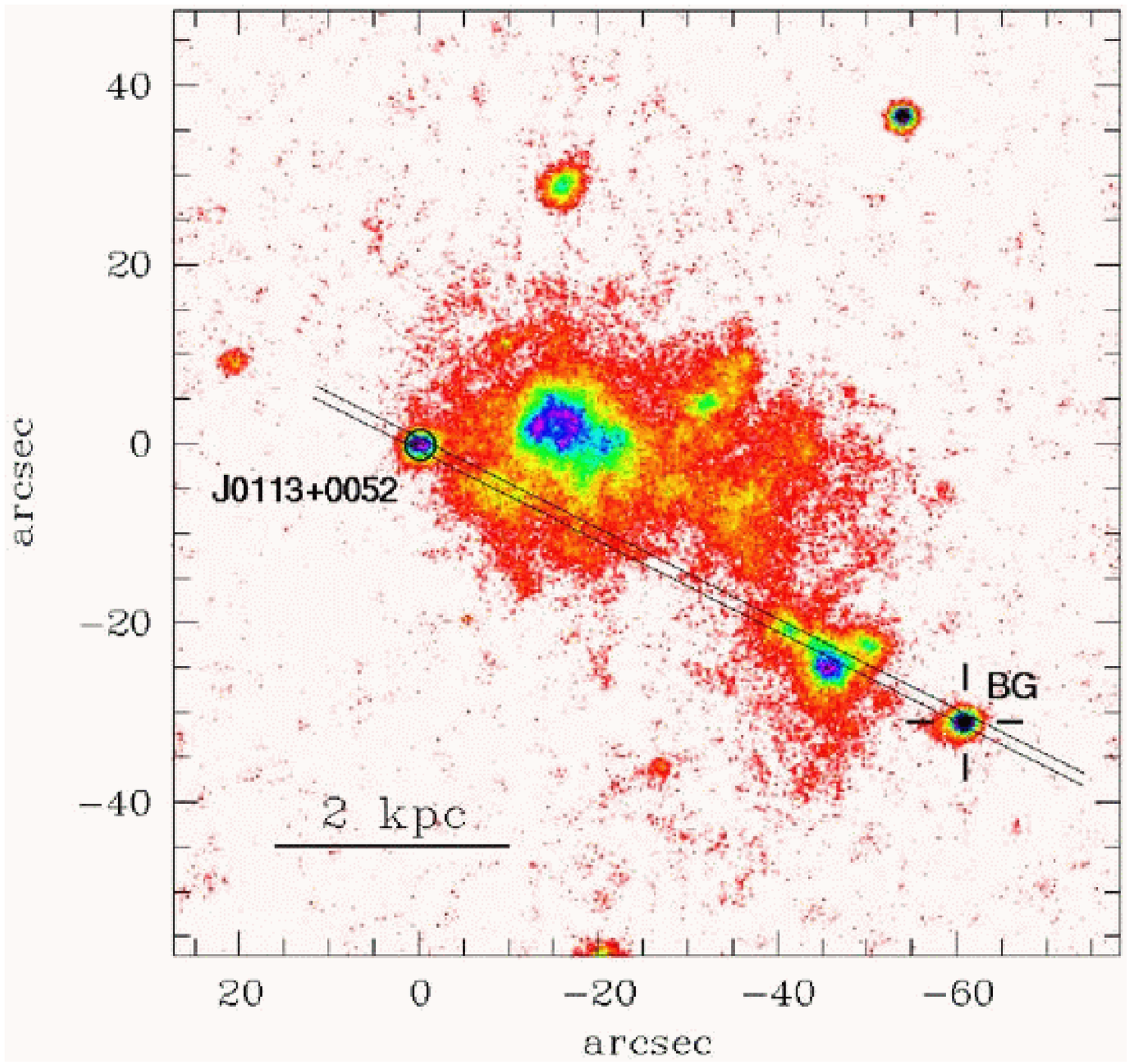,angle=0,width=8.0cm,clip=}
\caption{Acquisition images of J2104$-$0035 (left) and J0113$+$0052 
(right) in the $V$ band with the long-slit positions superposed. 
The overlaid circles with a diameter of 3\arcsec\ indicate the apertures 
out of which SDSS spectra were extracted. 
The linear scale depicted by the horizontal bar at the lower part 
of each panel was computed assuming a distance of 19.2 Mpc and 16.0 Mpc 
for J2104$-$0035 and J0113$+$0052, respectively. Distances 
were obtained from the observed redshift and adopting a Hubble constant of 
75 km s$^{-1}$ Mpc$^{-1}$. A bright star in the left panel and a background 
galaxy in the right panel are labeled respectively as ''star'' and ''BG''.
North is up, east is to the left.
}
\label{fig1}
\end{figure*}

\begin{table}
\caption{Coordinates of the Galaxies (J2000.0) \label{tab1}}
\begin{tabular}{lcc} \hline
Name       & R.A.   & DEC.   \\ 
\hline
J2104$-$0035& 21 04 55.3&$-00$ 35 22 \\
J0113$+$0052& 01 13 40.4&$+00$ 52 39 \\ \hline
\end{tabular}
\end{table}

Because of the scarcity of extremely low-metallicity galaxies such as I Zw 18 
and SBS 0335--052W, we stand a better
chance of finding them in very large spectroscopic surveys. 
One of the best surveys suitable for
such a search is the Sloan Digital Sky Survey (SDSS) \citep{Y00}. 
However, despite
intensive studies of galaxies with the definitely detected 
temperature-sensitive [O {\sc iii}]
$\lambda$4363 emission line in their spectra, no emission-line galaxy
with an oxygen abundance as low as that of I Zw 18 has been 
discovered in the SDSS Data Release 3 (DR3) and earlier releases. 
The lowest-metallicity emission-line galaxies found so far in these releases 
have oxygen abundances 12 + log O/H $>$ 7.4 \citep{K04a,K04b,I04,I06}. 

In order to find new candidates for extremely metal-deficient 
emission-line galaxies we carried out a systematic search 
for such objects in SDSS Data Release 4 (DR4), 
on the basis of the relative fluxes of emission lines. 
All extremely metal-deficient emission-line galaxies 
known are characterized by relatively weak (compared to 
H$\beta$) [O {\sc ii}] $\lambda$3727, [O {\sc iii}] $\lambda$4959, 
$\lambda$5007 and [N {\sc ii}] $\lambda$6583 emission 
lines \citep[e.g. ][]{IT98a,IT98b,I05,P05}.
These spectral properties select out uniquely low-metallicity 
dwarfs since no other type of galaxy possesses them. 
At variance with previous studies \citep{K04a,K04b,I04,I06}
we additionally considered spectra 
where the [O {\sc iii}] $\lambda$4363 emission line is weak or not detected.
Since the [O {\sc ii}] $\lambda$3727 line is out of the observed wavelength 
range 
in the SDSS spectra for all galaxies with a 
redshift $z$ lower than 0.02, we use the
[O {\sc iii}] $\lambda$4959, $\lambda$5007 and [N {\sc ii}] $\lambda$6583 
emission to pick out extremely low-metallicity galaxy candidates. 
Following this strategy we selected $\sim$ 20 galaxies with 
[O {\sc iii}]$\lambda$4959/H$\beta$ $\la$ 1 and 
[N {\sc ii}]$\lambda$6583/H$\beta$ $\la$ 0.05.

\section{Observations and Data Reduction}

We present here new spectroscopic observations of two galaxies,
J2104--0035 and J0113+0052 (Fig. \ref{fig1}), from the list of
extremely low-metallicity candidates selected from the SDSS. 
We use these new observations, together with SDSS spectra to show
that both galaxies have oxygen abundances nearly as low as those of the most 
metal-deficient emission-line galaxies known.
The coordinates of the two galaxies are shown in Table \ref{tab1}. 

\begin{figure}[t]
\hspace*{-0.0cm}\psfig{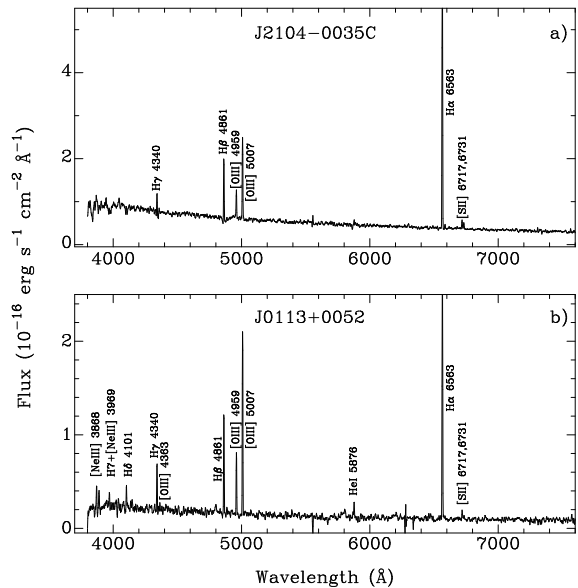}
\caption{SDSS spectra of J2104$-$0035C and J0113$+$0052.}
\label{fig2}
\end{figure}

Both galaxies have low redshifts, therefore the [O {\sc ii}] $\lambda$3727
emission line is blueward of the spectral range of the SDSS spectra
(Fig. \ref{fig2}). 
This fact, in conjunction with the non-detection of the weak 
[O {\sc ii}] $\lambda$7320,7331 emission lines in the SDSS spectra,
has prevented an oxygen abundance determination using publicly available SDSS data.

\begin{figure}
\hspace*{-0.0cm}\psfig{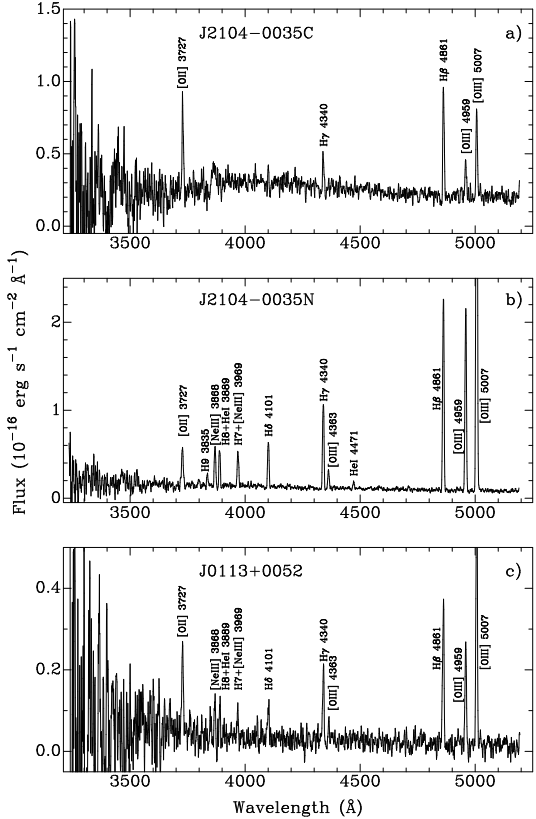}
\caption{3.6m ESO spectra of J2104$-$0035C, J2104$-$0035N and J0113$+$0052.}
\label{fig3}
\end{figure}

\begin{table*}
\caption{Emission Line Fluxes \label{tab2}}
\begin{tabular}{lcccccc} \hline
           & \multicolumn{2}{c}{SDSS}&& \multicolumn{3}{c}{3.6m} \\ \cline{2-3}
\cline{5-7}
Line       &J2104$-$0035C&J0113$+$0052&&J2104$-$0035C&
J2104$-$0035N&J0113$+$0052 \\ \hline
3727 [O {\sc ii}]              &      ...       &     ...       &&0.690$\pm$0.068&0.227$\pm$0.011&0.688$\pm$0.074 \\
3835 H9                        &      ...       &     ...       &&     ...       &0.082$\pm$0.012&      ...       \\
3869 [Ne {\sc iii}]            &      ...       &0.261$\pm$0.045&&     ...       &0.222$\pm$0.009&0.278$\pm$0.044 \\
3889 He {\sc i} + H8           &      ...       &0.190$\pm$0.057&&     ...       &0.217$\pm$0.010&0.297$\pm$0.044 \\
3968 [Ne {\sc iii}] + H7       &      ...       &0.163$\pm$0.057&&     ...       &0.201$\pm$0.010&0.178$\pm$0.044 \\
4101 H$\delta$                 &0.262$\pm$0.092 &0.252$\pm$0.058&&0.267$\pm$0.051&0.264$\pm$0.009&0.269$\pm$0.041 \\
4340 H$\gamma$                 &0.474$\pm$0.091 &0.487$\pm$0.077&&0.467$\pm$0.057&0.477$\pm$0.011&0.482$\pm$0.048 \\
4363 [O {\sc iii}]             &      ...       &0.075$\pm$0.035&&     ...       &0.097$\pm$0.005&0.086$\pm$0.016 \\
4471 He {\sc i}                &      ...       &     ...       &&     ...       &0.034$\pm$0.004&     ...        \\
4861 H$\beta$                  &1.000$\pm$0.107 &1.000$\pm$0.109&&1.000$\pm$0.058&1.000$\pm$0.018&1.000$\pm$0.056 \\
4959 [O {\sc iii}]             &0.405$\pm$0.064 &0.616$\pm$0.076&&0.334$\pm$0.033&0.951$\pm$0.017&0.710$\pm$0.048 \\
5007 [O {\sc iii}]             &1.143$\pm$0.113 &1.852$\pm$0.169&&0.962$\pm$0.048&2.876$\pm$0.046&2.045$\pm$0.090 \\
5876 He {\sc i}                &      ...       &0.175$\pm$0.046&&     ...       &     ...       &     ...        \\
6563 H$\alpha$                 &2.886$\pm$0.259 &2.617$\pm$0.255&&     ...       &     ...       &     ...        \\
6583 [N {\sc ii}]              &0.051$\pm$0.038 &0.033$\pm$0.032&&     ...       &     ...       &     ...        \\
6717 [S {\sc ii}]              &0.089$\pm$0.038 &      ...      &&     ...       &     ...       &     ...        \\
6731 [S {\sc ii}]              &0.066$\pm$0.034 &      ...      &&     ...       &     ...       &     ...        \\ 
$C$(H$\beta$)                   &      0.365     &     0.000     &&    0.025      &     0.000     &    0.000       \\
EW(H$\beta$)$^a$                   &      21        &     41        &&      42       &     203       &     193        \\
$F$(H$\beta$)$^b$                  &      9.7       &     6.4       &&     6.4       &    16.7       &     2.8        \\
EW(abs)$^a$                        &      1.2       &     0.0       &&     1.6       &     0.8       &     0.0        \\
\hline
\end{tabular}

$^a$ In \AA.

$^b$ In units of 10$^{-16}$ erg s$^{-1}$ cm$^{-2}$.
\end{table*}

\begin{table*}[t]
\caption{Element Abundances \label{tab3}}
\begin{tabular}{lcccc} \hline
           & \multicolumn{1}{c}{SDSS}&& \multicolumn{2}{c}{3.6m} \\ \cline{2-2}
\cline{4-5}
Property       &J0113$+$0052$^a$&&J2104$-$0035N&J0113$+$0052 \\ \hline
$T_e$(O {\sc iii}), K           &22520$\pm$7360 &&20060$\pm$620  &22880$\pm$2990  \\
$T_e$(O {\sc ii}), K            &15480$\pm$4510 &&15610$\pm$450  &15410$\pm$1780  \\
O$^+$/H$^+$, ($\times$10$^5$)     &0.567$\pm$0.421&&0.182$\pm$0.016&0.575$\pm$0.179 \\
O$^{2+}$/H$^+$, ($\times$10$^5$)  &0.836$\pm$0.593&&1.630$\pm$0.116&0.906$\pm$0.256 \\
O/H, ($\times$10$^5$)             &1.403$\pm$0.727&&1.812$\pm$0.117&1.481$\pm$0.313 \\
12+log O/H                        &7.15$\pm$0.23  &&7.26$\pm$0.03  &7.17$\pm$0.09   \\ 
N$^+$/H$^+$, ($\times$10$^7$)     &2.479$\pm$2.041&&     ...       &      ...       \\
$ICF$(N)                          &      2.49     &&     ...       &      ...       \\
N/H, ($\times$10$^7$)             &5.662$\pm$5.049&&     ...       &      ...       \\
log N/O                           &--1.39$\pm$0.45&&     ...       &      ...       \\ 
Ne$^{2+}$/H$^+$, ($\times$10$^6$) &2.542$\pm$1.649&&2.794$\pm$0.211&2.624$\pm$0.771 \\
$ICF$(Ne)                         &      1.17     &&     1.04      &     1.17       \\
Ne/H, ($\times$10$^6$)            &2.981$\pm$2.768&&2.914$\pm$0.235&3.058$\pm$1.260 \\
log Ne/O                          &--0.67$\pm$0.46&&--0.79$\pm$0.05&--0.69$\pm$0.20 \\
\hline
\end{tabular}

$^a$O$^+$/H$^+$ is obtained adopting the [O {\sc ii}] $\lambda$3727 emission
line flux from the 3.6m observations.
\end{table*}

New spectra of six
extremely low-metallicity candidates were obtained
on 7 and 8 October, 2005 with the  
EFOSC2 (ESO Faint Object Spectrograph and Camera) mounted at the 3.6m ESO 
telescope at La Silla.
Here we present the results for the two most metal-deficient emission-line 
galaxies J2104--0035 and J0113+0052 with 12 + log O/H $<$ 7.3. 
The spectroscopic properties of the remaining four emission-line 
galaxies with slightly higher oxygen abundance in the range 
12 + log O/H = 7.3 -- 7.6 will be discussed in a separate publication. 

The observing conditions were photometric during those two nights.
All 3.6m observations were performed with the same instrumental setup. 
We used the grism $\#07$ ($\lambda$$\lambda$3200--5200) with 
the grating 600 gr/mm. 
The long slit with a width of 1\farcs2 
was placed as it is shown in Fig. \ref{fig1}.
The spatial scale along the slit is of 0\farcs157 pixel$^{-1}$, and the 
spectral resolution is $\sim$6.2~\AA\ (FWHM).

Both galaxies were observed at low airmass $\la$ 1.2, therefore
no corrections for atmospheric refraction are needed.
The cometary galaxy J2104--0035 was observed with the slit oriented in the 
position angle P.A. = +16.5$^\circ$, so as to obtain simultaneously the spectra
of the central region of the galaxy (J2104--0035C) and the northern 
H {\sc ii} region (J2104--0035N). Note that only the J2104--0035C spectrum is
present in the SDSS DR4 and no public data are available for the northern
component, located at $\sim$ 6\arcsec\ from the central
part of the galaxy. The total exposure time was 5400 s. 
The galaxy J0113+0052 was observed
with the slit oriented in the position angle +64$^\circ$. The total exposure
time was 3600 s. 

The 3.6m data were reduced with the IRAF\footnote{IRAF is 
the Image Reduction and Analysis Facility distributed by the 
National Optical Astronomy Observatory, which is operated by the 
Association of Universities for Research in Astronomy (AURA) under 
cooperative agreement with the National Science Foundation (NSF).}
software package. This includes bias--subtraction, 
flat--field correction, cosmic-ray removal, wavelength calibration, 
night sky background subtraction, correction for atmospheric extinction and 
absolute flux calibration of the two--dimensional spectrum.
We extracted one-dimensional spectra of J2104--0035C and J2104--0035N 
respectively  
within apertures of 1\farcs2$\times$2\farcs5 and 1\farcs2$\times$0\farcs6,
and the one-dimensional
spectrum of J0113+0052 within a 1\farcs2$\times$0\farcs6 
aperture. The slit in 
Fig. \ref{fig1} (right panel) crosses several other knots with weak emission
lines in their spectra at the same redshift as those in J0113+0052 except for
the background galaxy labeled BG in Fig. \ref{fig1} (right panel).
The 3.6m and SDSS spectra were corrected for interstellar extinction 
using the reddening curve by \citet{W58} and for
redshift $z$, derived from the observed wavelengths of the emission lines. 
From the 3.6m spectra, we obtained $z$ = 0.00438 $\pm$ 0.00025 for
J2104--0035C, $z$ = 0.00479 $\pm$ 0.00014 for J2104--0035N and
$z$ = 0.00400 $\pm$ 0.00014 for J0113+0052. These values compare
well with the redshifts derived from the SDSS spectra: 
0.00468 for J2104--0035C and 0.00376 for J0113+0052.
Redshift-corrected one-dimensional SDSS and 3.6m spectra of the 
two emission-line galaxies are shown in Figs.~\ref{fig2} and \ref{fig3},
respectively.

Emission line fluxes in the 3.6m and SDSS spectra 
were measured using Gaussian profile fitting. 
The errors of the line fluxes in the 3.6m spectra 
were calculated from the photon statistics 
in the non-flux-calibrated spectra. As for SDSS spectra,
the errors of emission-line fluxes were obtained using the files with
errors in each pixel accompanying every SDSS spectrum.
The flux errors have been propagated in the 
calculations of the elemental abundance errors.

The observed fluxes of all hydrogen Balmer lines except for
the H7 and H8 lines were used to determine the interstellar extinction and 
the underlying stellar absorption. The two excluded lines are blended with other 
strong emission lines and could not therefore be used for the determination
of interstellar reddening. The extinction coefficient $C$(H$\beta$) and 
the equivalent 
width of absorption hydrogen lines EW$_{abs}$ are derived by 
minimizing deviations of corrected hydrogen emission line fluxes from the 
theoretical recombination values. In this procedure we assumed that
EW$_{abs}$ is the same for all hydrogen lines.
The corrected emission line fluxes $I$($\lambda$) relative to the H$\beta$  
fluxes, the extinction coefficients $C$(H$\beta$), the 
equivalent widths EW(H$\beta$),
the observed H$\beta$ fluxes $F$(H$\beta$), and the 
equivalent widths of the hydrogen absorption lines 
are listed in Table \ref{tab2} both for 3.6m and SDSS spectra. 

\section{Results}

The electron temperature $T_{\rm e}$, ionic and total heavy element abundances 
were derived following \citet{I06}. In particular, for the 
O$^{2+}$ and Ne$^{2+}$ ions we adopt
the temperature $T_e$(O {\sc iii}) derived from the 
[O {\sc iii}] $\lambda$4363/($\lambda$4959 + $\lambda$5007)
emission line ratio. 
The N$^+$ and O$^+$ abundances were derived with the temperature
$T_e$(O {\sc ii}). The latter was obtained from the relation between $T_e$(O {\sc iii})
and $T_e$(O {\sc ii}) of \citep{I06}.

The [O {\sc iii}] $\lambda$4363 emission line was not detected in
J2104--0035C. Therefore, no abundance determination was possible for this
region. The [O {\sc ii}] $\lambda$3727 line is beyond the 
wavelength range of the SDSS spectrum of J0113+0052 while 
[O {\sc ii}] $\lambda$7320, 7331 emission lines are too weak.
In principle, the flux of the former
could be estimated from the [N {\sc ii}] $\lambda$6583 emission line 
assuming the N/O abundance ratio typical for the low-metallicity galaxies. 
However, the [N {\sc ii}] $\lambda$6583 emission line is weak and its flux is derived
with a large error (Table \ref{tab2}). On the other hand,
the fluxes relative to the H$\beta$ of common emission lines in the 
3.6m and SDSS spectra of J0113+0052 
are similar despite the different apertures used in both observations.
Therefore, to derive the oxygen abundance of this galaxy 
from the SDSS spectrum,
we have adopted the relative flux [O {\sc ii}] $\lambda$3727/H$\beta$ 
determined from the 3.6m spectrum. 

The 3.6m spectra covered the blue wavelength region,
so that the [S {\sc ii}] $\lambda$6717, 6731 emission lines, usually 
used for the determination
of the electron number density, were not observed. 
In the SDSS spectra, these lines
are weak in both galaxies. Therefore for abundance determinations, we have 
adopted  
$N_e$ = 100 cm$^{-3}$. The precise value of the electron number density 
makes little difference in the derived abundances
since in the low-density limit ($N_e$ $\la$ 10$^3$ cm$^{-3}$) 
which holds for the H {\sc ii} regions
considered here, the element abundances do not depend sensitively 
on $N_e$.

The electron temperatures $T_{\rm e}$(O {\sc iii}) and $T_{\rm e}$(O {\sc ii}) 
for the high and low-ionization zones in H {\sc ii} regions respectively,
the ionization correction factors ($ICF$s) and the 
ionic and total heavy element abundances for 
oxygen, neon, and nitrogen are shown in Table \ref{tab3} for J2104--0035N 
(3.6m data) and for J0113+0052 (SDSS and 3.6m data).

The oxygen abundance in J2104--0035N
is 12 + log O/H = 7.26 $\pm$ 0.03, slightly higher than in I Zw 18, but 
lower than the abundance 12 + log O/H = 7.31 $\pm$ 0.01 of the 
brightest part of the
BCD SBS 0335--052E \citep{TI05}. 
Thus, J2104--0035 belongs to the very rare emission-line
galaxies with an oxygen abundance below 7.3 \citep{SS70,I05,P05}. 
No galaxy with comparable metallicity has been found in previous 
studies of the SDSS by \citet{K04a,K04b} and \citet{I04,I06}. 
The oxygen abundance derived for J0113+0052 is more uncertain because of its 
noisier 
spectra. However, the abundances 
12 + log O/H = 7.15 $\pm$ 0.23 and 7.17 $\pm$ 0.09 derived independently 
from the SDSS and 3.6m spectra, are consistent and suggest
that this galaxy is also extremely metal-poor. 
Its oxygen abundance compares well with
that of the northwest component of I Zw 18. 
However, new higher signal-to-noise observations
are necessary to draw more definite conclusions 
about the metallicity of this galaxy.
The N/O and Ne/O abundance ratios in the two galaxies
are consistent within the errors with the ratios
obtained for other extremely metal-deficient galaxies \citep{IT99}.

\section{Summary}

We present 3.6m ESO telescope spectroscopic observations of the two extremely 
metal-deficient galaxies J2104--0035 and J0113+0052 selected from the Data 
Release 4 (DR4) of the Sloan Digital Sky Survey (SDSS). The oxygen abundances 
of J2104--0035 and J0113+0052 are 12 +log O/H = 7.26 $\pm$ 0.03 and 
7.17 $\pm$ 0.09, respectively, making them the most metal-deficient 
emission-line galaxies found thus far in the SDSS. The two galaxies belong to 
the same very scarce class of BCDs with 12 +log O/H below 7.3 consisting 
of SBS 0335--052W \citep[12 + log O/H = 7.12 $\pm$ 0.03, ][]{I05}, 
I Zw 18 \citep[12 + log O/H = 7.17 $\pm$ 0.01, ][]{TI05} and DDO 68
\citep[12 + log O/H = 7.21 $\pm$ 0.03, ][]{P05}. In the future we plan to 
carry out spectroscopic observations for the whole sample of $\sim$ 20 
candidates selected from the SDSS DR4. We hope to increase
the number of the extremely low-metallicity galaxies and to put limits on
their spatial density in the local universe.

\begin{acknowledgements}
Y. I. I. and N. G. G. thank the hospitality of the Institute for Astrophysics
(G\"ottingen), the support of the DFG grant No. 436 UKR 17/25/05 and of the 
grant No 02.07.00132 from the  Ukrainian Fund of Fundamental Investigations.
P.P. would like to thank Gaspare Lo Curto, Lorenzo Monaco, Carlos La Fuente, 
Eduardo Matamoros and the whole ESO staff at the La Silla Observatory for their support.
Y. I. I. and T. X. T. acknowledge the partial financial support of NSF
grant AST 02-05785. T. X. T. thanks the hospitality of the Institut 
d'Astrophysique in Paris and of the Service d'Astrophysique at Saclay during 
his sabbatical leave. He is grateful for a Sesquicentennial Fellowship 
from the University of Virginia.  
The research described in this publication was made
possible in part by Award No. UP1-2551-KV-03 of the US Civilian Research
\& Development Foundation for the Independent States of the Former
Soviet Union (CRDF).
All the authors acknowledge the work of the Sloan Digital Sky
Survey (SDSS) team.
Funding for the SDSS has been provided by the
Alfred P. Sloan Foundation, the Participating Institutions, the National
Aeronautics and Space Administration, the National Science Foundation, the
U.S. Department of Energy, the Japanese Monbukagakusho, and the Max Planck
Society. The SDSS Web site is http://www.sdss.org/.
\end{acknowledgements}

\end{document}